\documentclass[journal=jctcce,manuscript=article]{achemso}
\usepackage[version=3]{mhchem} % Formula subscripts using \ce{}
\usepackage[T1]{fontenc}       % Use modern font encodings
\usepackage{amsmath}
\usepackage{subfigure}
\usepackage{booktabs}
\usepackage{psfrag}
\usepackage{xspace}
\usepackage{placeins}
\setkeys{acs}{articletitle = true}

\author{John W.\ R.\ Morgan}
\email{jwrm2@cam.ac.uk}
\affiliation{University Chemical Laboratories, Lensfield Road, Cambridge CB2 1EW, UK}

\author{Sharon C.\ Glotzer}
\affiliation{Department of Chemical Engineering, University of Michigan, Ann Arbor, USA}
\alsoaffiliation{Department of Materials Science and Engineering, University of Michigan, Ann Arbor, USA}
\alsoaffiliation{Biointerfaces Institute, University of Michigan, Ann Arbor, USA,}

\title[OPP Clusters]{Energy Landscapes for Digital Alchemy}

\begin{document}

\begin{abstract}
We apply energy landscape methods to digital alchemy, defining a system
in which the parameters of the potential are
treated as degrees of freedom. Using geometrical optimisation, we locate minima and
transition states on the landscape for small clusters.
We show that it is easy to find the
parameters that give the lowest energy minimum, and that the distribution of minima on
the alchemical landscape is concentrated in particular areas. We also conclude that
the alchemical landscape is more frustrated, in terms of competition between low
energy structures separated by high barriers. Transition states on the
alchemical landscape are classified by whether they become minima or transition
states when the potential parameters are fixed. Those that become minima have
a significant alchemical component, while those that remain as transition states
can be characterised mainly in terms of atomic displacements.
\end{abstract}

\section{Introduction}
A fundamental challenge in colloidal science is posed by the vast array of possible
colloids that can be made.\cite{ZhangG04, GlotzerS07, PawarK10, SacannaP11} The
complexity and time required to physically manufacture designer particles means
that is not possible to fully explore the parameter space. Computer simulations
allow a greater throughput and there has been some success using the energy
landscape approach\cite{EnergyLandscapes} to predict assembly pathways and structures
for colloidal systems.\cite{MorganCDW13, MorphewC16, FejerW15}

Typically, a computational investigation involves defining a potential with a
set of parameters to describe the system. Simulations may be run at different
values of the parameters to explore the behaviour or locate a region of parameter
space supporting behaviour of interest. An alternative approach is to allow the
parameters themselves to vary. This
technique is known as digital alchemy, as it involves changing the nature of the particles
modelled. The model parameters are already included in
the potential energy function. These parameters become variables rather than
fixed quantities and can be changed by step taking algorithms in Monte Carlo simulations or evolved dynamically
in molecular dynamics simulations, in a thermodynamically consistent way.\cite{vanAndersKSDG15}
Previous work has mostly focused on changing the particle shape, but the approach
can be applied much more generally. By exploring a landscape with such extended degrees
of freedom, it may be possible to locate the optimal parameters
for a particular structure much more quickly and to identify features that favour 
a particular target.
\cite{vanAndersKSDG15, DuANG17, CersonskyADG18}

Following previous work\cite{vanAndersKSDG15} we investigate the 
potential energy landscape of the oscillating pair potential (OPP).\cite{MihalkovicH12}
Originally developed in the context of metal alloys, this potential displays interesting
phase behaviour, including an icosahedral quasicrystal.\cite{EngelDPG15}
Here we exploit it as a toy potential with two adjustable parameters that displays
many local minima in the alchemical landscape. We define the alchemical landscape to
mean the potential energy landscape including both the particle Cartesian
coordinates and the alchemical coordinates as variables. As proof of concept, we survey
the alchemical landscape for small clusters of
particles, locating energy minima and transition states, and compare
with fixed parameter landscapes in various ranges. We also
make some observations on the nature of transition states on the alchemical
landscape.

\section{Methods}
\subsection{Oscillating Pair Potential}
The form of the OPP is
\begin{align}
V\left(\mathbf{r}\right) =& \sum_{i \neq j} V_{ij} \left(\mathbf{r}_{i}, \mathbf{r}_{j}\right), \nonumber \\
V_{ij} \left(\mathbf{r}_{i}, \mathbf{r}_{j}\right) =& \frac{\displaystyle \epsilon \sigma^{15}}{\displaystyle r_{ij}^{15}} +
\frac{\displaystyle \epsilon \sigma^{3}}{\displaystyle r_{ij}^{3}} \cos \left(k \left(\frac{r_{ij}}{\sigma} - 1.25\right) - \phi\right),
\end{align}
where $r_{ij}$ is the distance between a pair of particles and the sum runs over all pairs
of particles. $V_{ij} \left(\mathbf{r}_{i}, \mathbf{r}_{j}\right)$ is the pair
potential, $\epsilon$ is the unit of energy and $\sigma$ is the unit of distance.
The alchemical parameters are $k$ and $\phi$. $k$ represents a frequency and
controls the spacing between wells: a higher value of $k$ means there are more wells in the
potential. $k$ was restricted to the range $5 < k < 20$ using a harmonic repulsion away from
forbidden regions

\begin{equation}
V_k\left(k\right) = \left\lbrace
\begin{array}{ll}
    \epsilon k_{rep} \left(k - 5\right)^{2}, & \text{if}~k < 5, \\ 
    \epsilon k_{rep} \left(k - 20\right)^{2}, & \text{if}~k > 20, \\
    0, & \text{otherwise}.
\end{array} \right.
\end{equation}

A value of $k_{rep} = 10^{4}$ was sufficient for the harmonic repulsion constant. Any minima
or transition states located outside the acceptable $k$ range were ignored.
$\phi$ represents a phase. $\phi$ is periodic over
the range $0 \leq \phi < 2\pi$, so no constraints were necessary.

Usually, a cutoff is not required in the potential when modelling clusters. Here, however,
the potential continues to oscillate even at very long range, so without a cutoff there
are actually an infinite number of wells and an infinite number of local potential energy minima.
To avoid this scenario, a cutoff was introduced. The Stoddard-Ford procedure\cite{StoddardF73} proved
to be unsuitable, as the continuing oscillations could make the gradient 
far from zero at the cutoff distance, introducing very large changes to the potential.
Instead the XPLOR\cite{AndersonLT08, GlaserNALSMMG15}
smoothing function was used, with the smoothing beginning at $r_{ij} = 2.5$
and the potential and gradient going smoothly to $0$ at $r_{ij} = 3$.

\subsection{Generating Databases}
To generate a database of minima and transition states, basin-hopping (BH)
global optimisation\cite{EnergyLandscapes, LiS87, LiS88, WalesD97}
as implemented in \textsf{GMIN}\cite{GMIN} was first run to locate the
global potential energy minimum and other low-lying minima. These minima were used as
end points for transition state searches using the 
doubly-nudged\cite{TrygubenkoW04, TrygubenkoW04a, SheppardTH08}
elastic band\cite{MillsJ94, MillsJS95, NEB, HenkelmanUJ00, HenkelmanJ00} algorithm
and eigenvector following\cite{MunroW99} as implemented in \textsf{OPTIM}\cite{OPTIM}.
A \textsf{PATHSAMPLE}\cite{PATHSAMPLE} database was initialised from these connections and expanded
using additional double-ended and single-ended\cite{MunroW99, WalesDMMW00, KumedaMW01}
transition state searches. The resulting database constitutes a kinetic transition network,
from which phenomenological rates and thermodynamic properties can be extracted.\cite{Wales18}

\subsection{Digital Alchemy}
During alchemical simulations, the potential parameters $k$ and $\phi$ are allowed to
vary and are optimised in the same manner as particle coordinates. The BH and
transition state search
procedures do not depend on the nature of the variables being optimised, so no
modifications were needed. The BH procedure requires the first derivatives of the
pair potential with respect to the parameters, which are
\begin{align}
\frac{\displaystyle \partial V_{ij}}{\displaystyle \partial k} =&
- \epsilon \sin \left(k \left(\frac{r_{ij}}{\sigma} - 1\right) - \phi \right) \left(\frac{r_{ij}}{\sigma} - 1\right) \frac{\sigma^{3}}{r_{ij}^{3}}, \nonumber \\
\frac{\displaystyle \partial V_{ij}}{\displaystyle \partial \phi} =&
\epsilon \sin \left(k \left (\frac{r_{ij}}{\sigma} - 1\right) - \phi \right) \frac{\sigma^{3}}{r_{ij}^{3}}.
\end{align}
Normal mode analysis and 
some transition state search algorithms benefit from analytical second derivatives, 
which include the mixed
derivatives with respect to position and potential parameters. These derivatives
are also straightforward; the explicit results are omitted for brevity.
Appropriate products with the XPLOR smoothing
function and its derivatives must also be included.

The potential energy surface does not depend on the temperature or the mass of the
particles. Hence questions about what mass is associated with the
alchemical degrees of freedom and whether their response to temperature is the
same as for particle Cartesian degrees of freedom are avoided. However, for calculating free
energies, these questions must be addressed. In this work, the main results do
not depend on this choice, but we also calculate heat
capacities as a post-processing tool only, using the same mass and
temperature for the alchemical degrees of freedom as for the atomic coordinates.

\subsection{Frustration Index}

A frustration index can be used to quantify the amount of frustration present in a landscape,
\cite{BryngelsonOSW95,OnuchicLW97} defined in terms of low energy minima separated
by high barriers. Here we use a recently developed index based on the
barrier heights between the global minimum and the other minima.\cite{deSouzaSNFW17} The definition
is

\begin{equation}
\widetilde{f}\left(T\right) = \sum_{\alpha \neq {\rm gmin}} 
\frac{p_{\alpha}^{\rm eq} \left(T\right)}{1 - p_{\rm gmin}^{\rm eq} \left(T\right)}
\left(
    \frac{V_{\alpha}^{\dag} - V_{\rm gmin}}{V_{\alpha} - V_{\rm gmin}}
\right),
\end{equation}
where the sum runs over all minima other than the global minimum, $p^{\rm eq}_{\alpha}$ is
the equilibrium occupation probability of minimum $\alpha$, $V_{\alpha}$ is the
potential energy of minimum $\alpha$, and $V_{\alpha}^{\dag}$ is the potential
energy of the highest transition state on the lowest energy path between
minimum $\alpha$ and the global minimum.

\section{Results}
We sought the smallest cluster of particles that would display interesting
behaviour when the alchemical landscape was explored. Four particles can all
be nearest neighbours (in a regular tetrahedral arrangement),
and the alchemical parameters then adjust to make the nearest neighbour well as deep
as possible. There are only a very few minima, corresponding to the $k$ and $\phi$
combinations that give a minimum in the well depth with respect to changing $k$ and $\phi$.

Five particles cannot all be nearest neighbours. Non-nearest neighbour interactions
can still be optimised when they are at a
distance corresponding to a different well in the OPP potential. With alchemical variation
of $k$ and $\phi$, these potential parameters can also adjust to make a particular
distance more favourable, leading to more complicated behaviour than for the four
particle cluster.

A database of minima and transition states was created for the five particle cluster,
including the alchemical degrees of freedom. Table \ref{tab:databasesizes} shows the
size of the database and the energy of the lowest minimum found. The database is not
complete, as evidenced by the absence of pathways between all pairs of minima, but
the difficulty of discovering new minima and transition states suggests that the majority
of structures have been located. In fact, the only pathway between a pair of minima
may consist of transition states and intermediate minima that lie outside the
permitted $k$ range, so even with a complete database, the network may not be fully
connected.

\begin{table}
    \begin{tabular}{ccccccc}
        $k$ & $\phi$ & $N_{\rm min}$ & $N_{\rm ts}$ & $N_{\rm con}$ & $E_{\rm min} / \epsilon$ & description\\ \hline
        \multicolumn{2}{c}{alchemical} & 6760 & 10386 & 2492 & $-4.9693$ \\
         6.00 & 1.00 &     61 &    714 &     58 & $-1.9792$ & low density \\
         8.74 & 4.37 &   2931 &  12691 &   1184 & $-4.9693$ & global minimum \\
        19.75 & 5.00 & 458977 & 489497 & 221179 & $-4.6385$ & high density
    \end{tabular}
    \caption{$k$ and $\phi$ values for selected databases, with the number of
             minima and transition states in those databases, the number of
             minima in the largest connected component, and the
             energy of the lowest minimum found. The reason for the selection
             of these $k$ and $\phi$ values is summarised in the final column.}
    \label{tab:databasesizes}
\end{table}

\begin{figure}
    \psfrag{ 20}{20}
    \psfrag{ 18}{18}
    \psfrag{ 16}{16}
    \psfrag{ 14}{14}
    \psfrag{ 12}{12}
    \psfrag{ 10}{10}
    \psfrag{ 8}{8}
    \psfrag{ 6}{6}
    \psfrag{k}{$k$}
    \psfrag{ 0}{0}
    \psfrag{ 1}{1}
    \psfrag{ 2}{2}
    \psfrag{ 3}{3}
    \psfrag{ 4}{4}
    \psfrag{ 5}{5}
    \psfrag{ 6}{6}
    \psfrag{phi}{$\phi$}
    \psfrag{Minima Density}{Minima Density}
    \psfrag{0}{0}
    \psfrag{3000}{3000}
    \psfrag{glob}{global minimum}
    \psfrag{great}{high density}
    \psfrag{low}{low density}
    \includegraphics{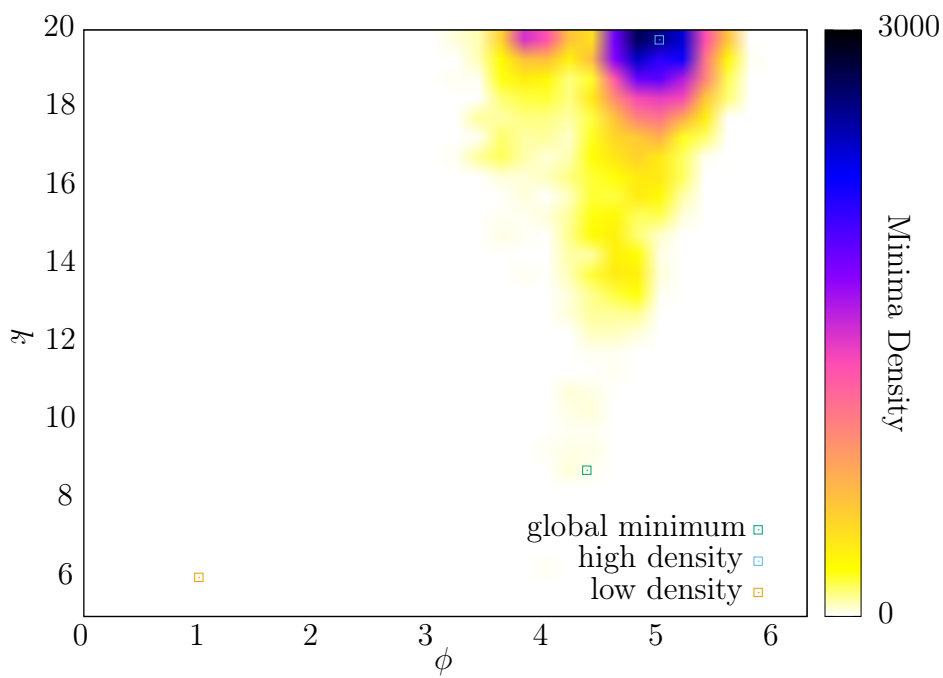}
    \caption{The density of minima as a function of the alchemical parameters
             $k$ and $\phi$. Most of the space has no minima at all. Unit density
             corresponds to one minimum in an area of one $k$ unit by one $\phi$ unit.
             The points chosen for generating fixed parameter databases are also shown.}
    \label{fig:minimumdensity}
\end{figure}

The distribution of minima throughout $k$ and $\phi$ space is shown in figure
\ref{fig:minimumdensity}. It is striking that most minima occur within a small
region of alchemical space, with most combinations of parameters having
no minima at all. Most minima occur around the $\phi$ value that gives the deepest
nearest-neighbour well and at higher $k$ values, meaning there are more wells in
the potential and more possible distances between pairs of particles.

\begin{figure}
    \psfrag{dist}{$r_{ij}$}
    \psfrag{pot}{$V_{ij} / \epsilon$}
    \psfrag{global minimum}{global minimum}
    \psfrag{high density}{high density}
    \psfrag{low density}{low density}
    \psfrag{-0.5}{$-0.5$}
    \psfrag{0}{$0.0$}
    \psfrag{0.5}{$0.5$}
    \psfrag{1}{$1.0$}
    \psfrag{1.5}{$1.5$}
    \psfrag{2}{$2.0$}
    \psfrag{2.5}{$2.5$}
    \psfrag{3}{$3.0$}
    \psfrag{ 1}{$1.0$}
    \psfrag{ 1.5}{$1.5$}
    \psfrag{ 2}{$2.0$}
    \psfrag{ 2.5}{$2.5$}
    \psfrag{ 3}{$3.0$}
    \includegraphics{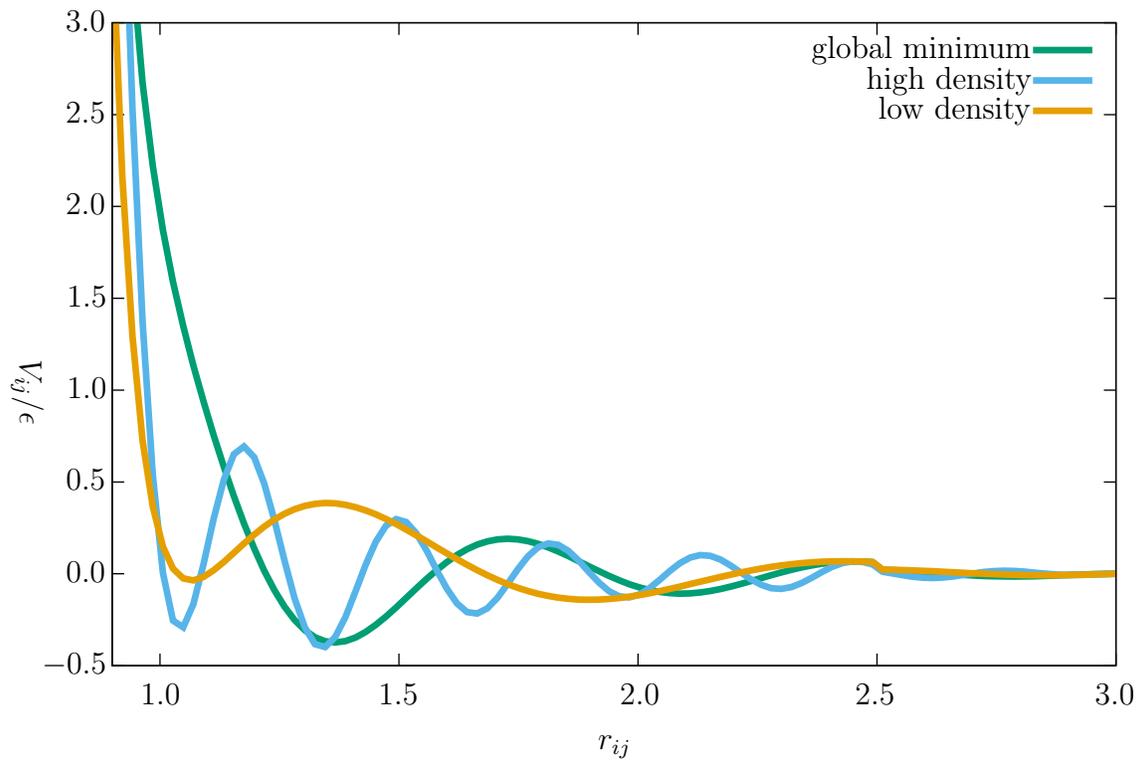}
    \caption{The OPP potential for the three sets of fixed parameters used, as listed in table \ref{tab:databasesizes}.}
    \label{fig:potential}
\end{figure}

Databases of minima and transition states at fixed alchemical parameters were also generated.
A landscape with fixed alchemical parameters is a slice through the full alchemical
landscape. Their parameters and
characteristics are summarised in table \ref{tab:databasesizes}, with plots of the potential in 
figure \ref{fig:potential}. Three alchemical points were selected,
corresponding to the global minimum from the alchemical database, a region with a low
density of minima, and a region with a high density of minima. In all cases, the
global minimum was a trigonal bipyramid (point group $D_{3h}$), differing only in
the distances between particles, which adjust to match the distance for the deepest
well in the potential.

The fixed parameter database in the low density region is unsurprisingly small,
and we note that there are no alchemical minima at these parameter values. At
high alchemical minima density, the number of minima with fixed parameters
is much larger than the total number of alchemical minima. Allowing alchemical
variation can reduce the number of minima and transition states on the landscape,
as most structures that are stationary points with the parameters fixed are not
stationary points when they are allowed to vary.

This result stands in
contrast to the usual behaviour when adding more degrees of freedom by adding more
particles, where the number of minima tends to increase exponentially on
increasing the number of particles.\cite{StillingerW84,WalesD03,MorganW14} Most minima at
fixed parameter values will not be minima in the extended space, as changes to
the potential can usually be made that decrease the energy. Conversely, on adding
a new particle to a system, minima will often exist with very similar pair distances
to structures before the addition. The main difference, then, is that changing
the alchemical parameters affects all the pair distances together.

\begin{figure}
    \begin{minipage}{.4\textwidth}
        \psfrag{a}{(a)}
        \psfrag{epsilon}{$\epsilon$}
        \includegraphics[width=0.95\textwidth]{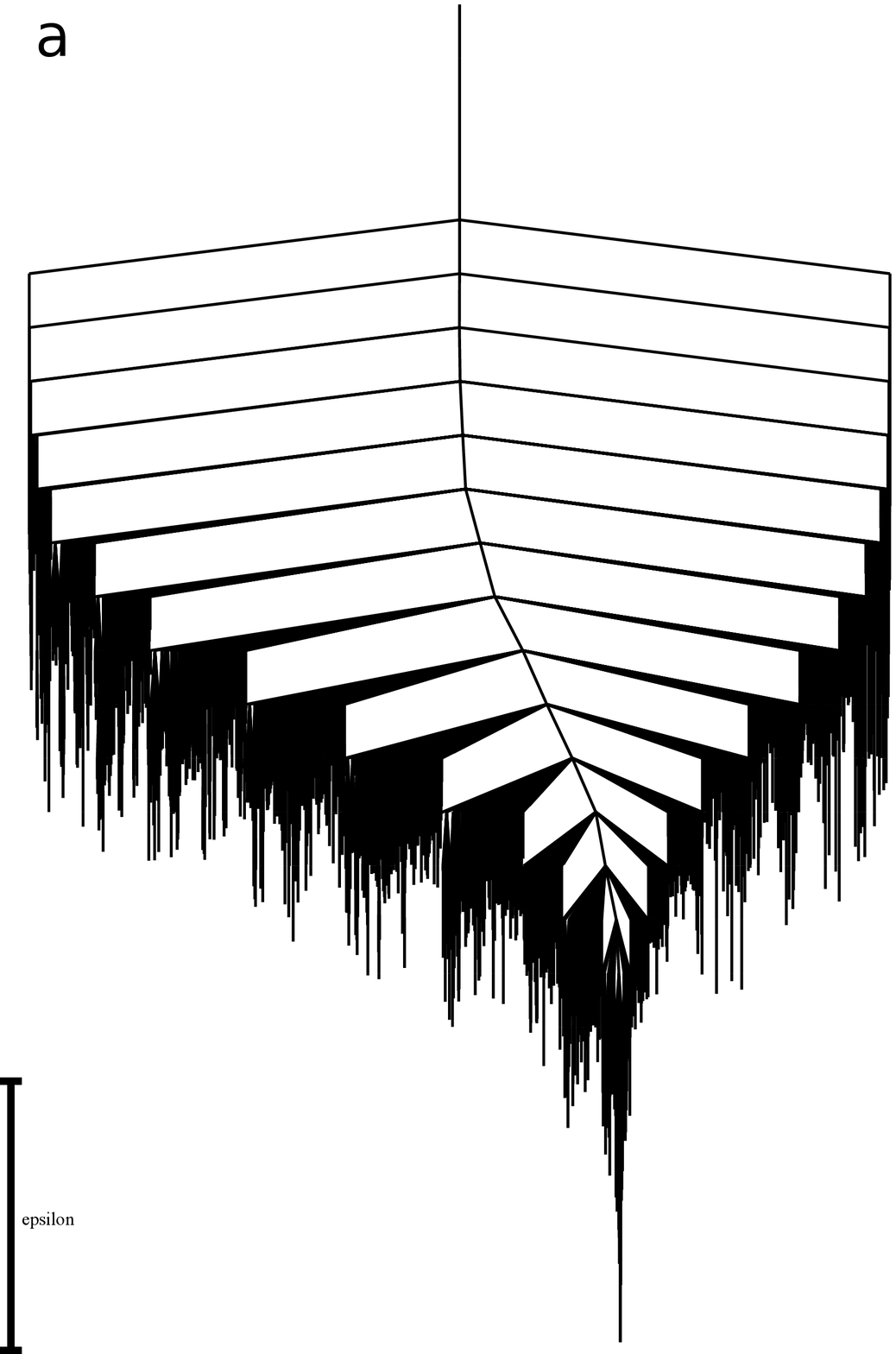}
    \end{minipage}
    \hspace{0.5cm}
    \begin{minipage}{.4\textwidth}
        \psfrag{b}{(b)}
        \psfrag{epsilon}{$\epsilon$}
        \includegraphics[width=0.95\textwidth]{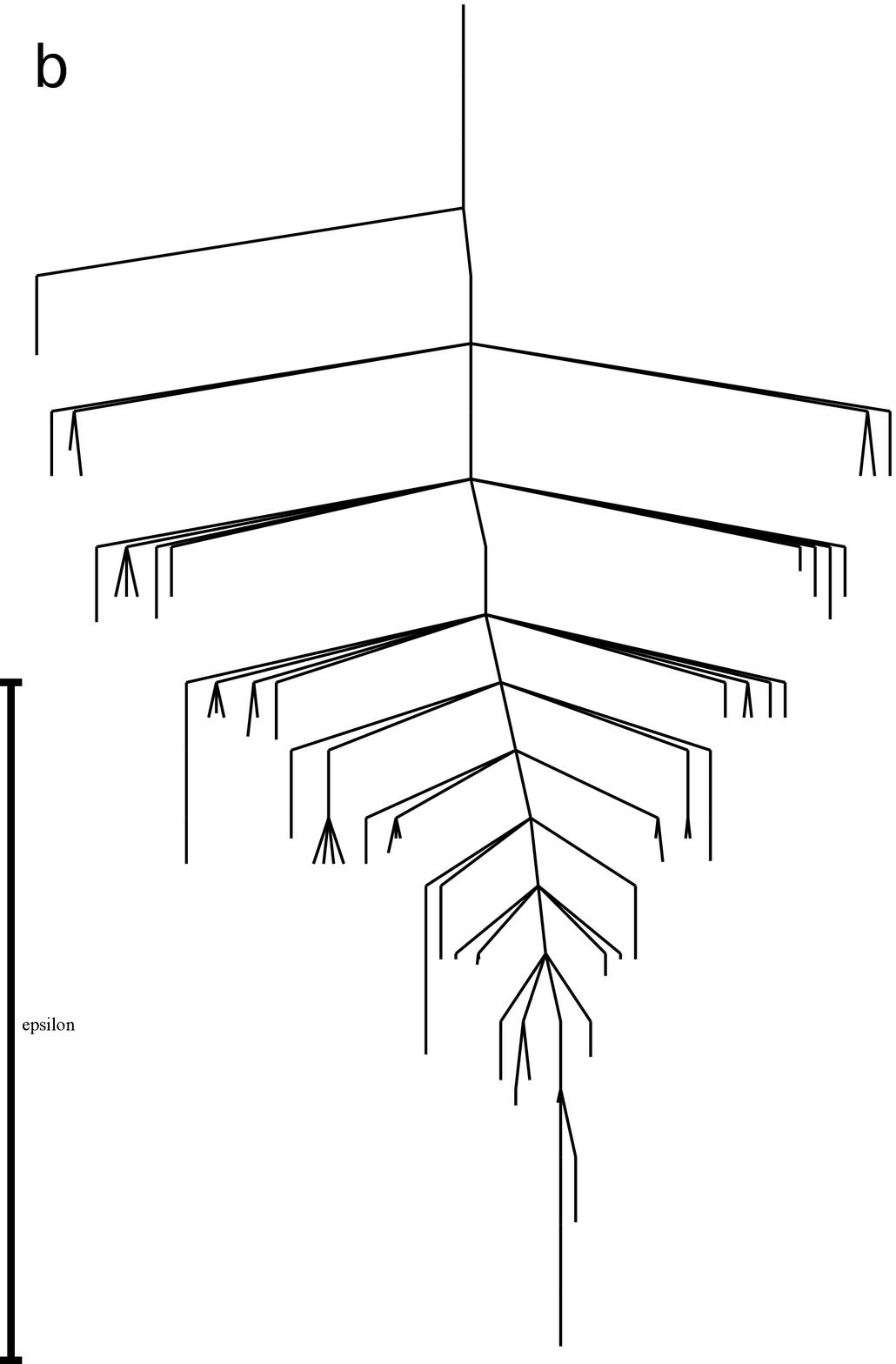}
    \end{minipage}
    \newline
    \begin{minipage}{.4\textwidth}
        \psfrag{c}{(c)}
        \psfrag{epsilon}{$\epsilon$}
        \includegraphics[width=0.95\textwidth]{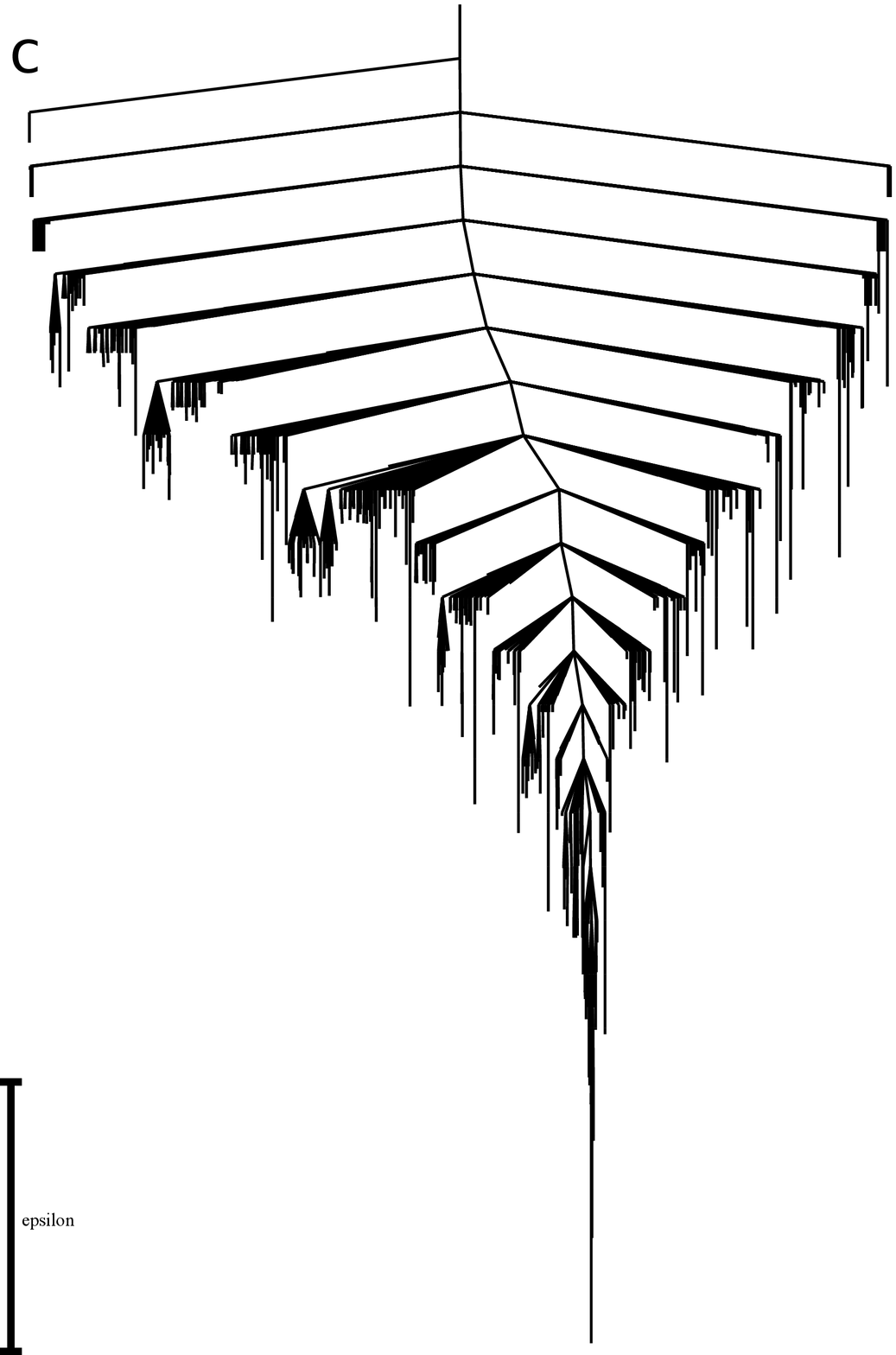}
    \end{minipage}
    \hspace{0.5cm}
    \begin{minipage}{.4\textwidth}
        \psfrag{d}{(d)}
        \psfrag{epsilon}{$\epsilon$}
        \includegraphics[width=0.95\textwidth]{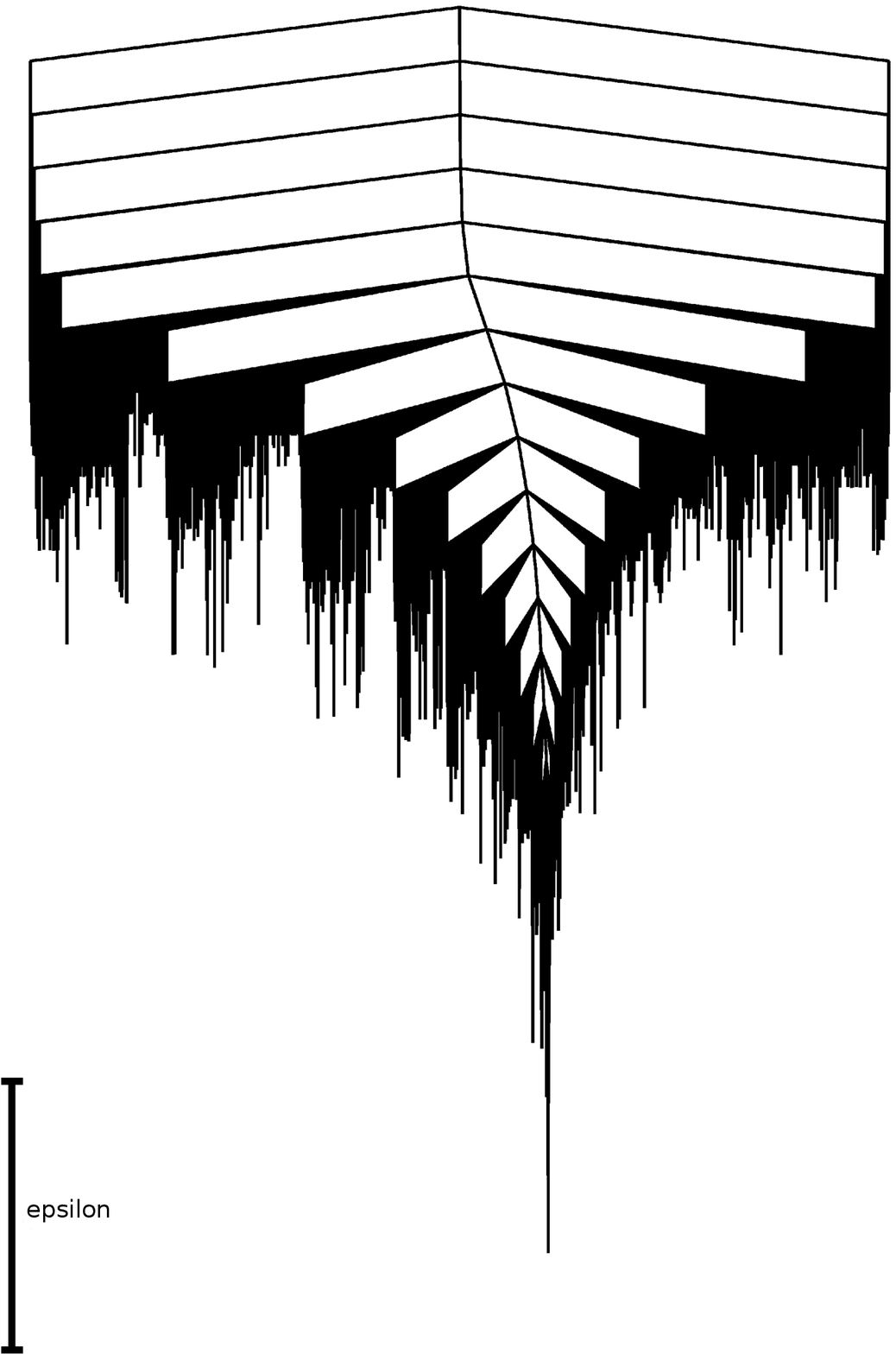}
    \end{minipage}
    \caption{Disconnectivity graphs for the alchemical landscape and the three fixed parameter landscapes, as listed in table \ref{tab:databasesizes},
        showing the connected component including the global minimum.
        (a) The alchemical system. (b) The low density system. (c) The global minimum system.
        (d) The high density system. }
    \label{fig:disconn}
\end{figure}

The landscapes for the four systems are summarised in the disconnectivity graphs
shown in figure \ref{fig:disconn}.\cite{BeckerK97,WalesMW98} Apart from the very
different numbers of minima, the structures of the graphs are fairly similar,
corresponding to ``palm tree'' funnelled landscapes.\cite{WalesMW98} The alchemical
system is most similar to the high density system. Taken in conjunction with the
density distribution in figure \ref{fig:minimumdensity}, it is clear that the
contributions from the high density region are dominating alchemical space.

\begin{figure}
    \psfrag{alch}{alchemical}
    \psfrag{glob}{global minimum}
    \psfrag{great}{high density}
    \psfrag{low}{low density}
    \psfrag{T}{$T/T_{m}$}
    \psfrag{f}{$\widetilde{f}$}
    \psfrag{ 10000}{10000}
    \psfrag{ 1000}{1000}
    \psfrag{ 100}{100}
    \psfrag{ 10}{10}
    \psfrag{ 1}{1}
    \psfrag{ 0.1}{0.1}
    \psfrag{ 0.01}{0.01}
    \psfrag{ 0}{0}
    \psfrag{ 1}{1}
    \psfrag{ 2}{2}
    \psfrag{ 3}{3}
    \psfrag{ 4}{4}
    \psfrag{ 5}{5}
    \psfrag{ 6}{6}
    \psfrag{ 7}{7}
    \includegraphics{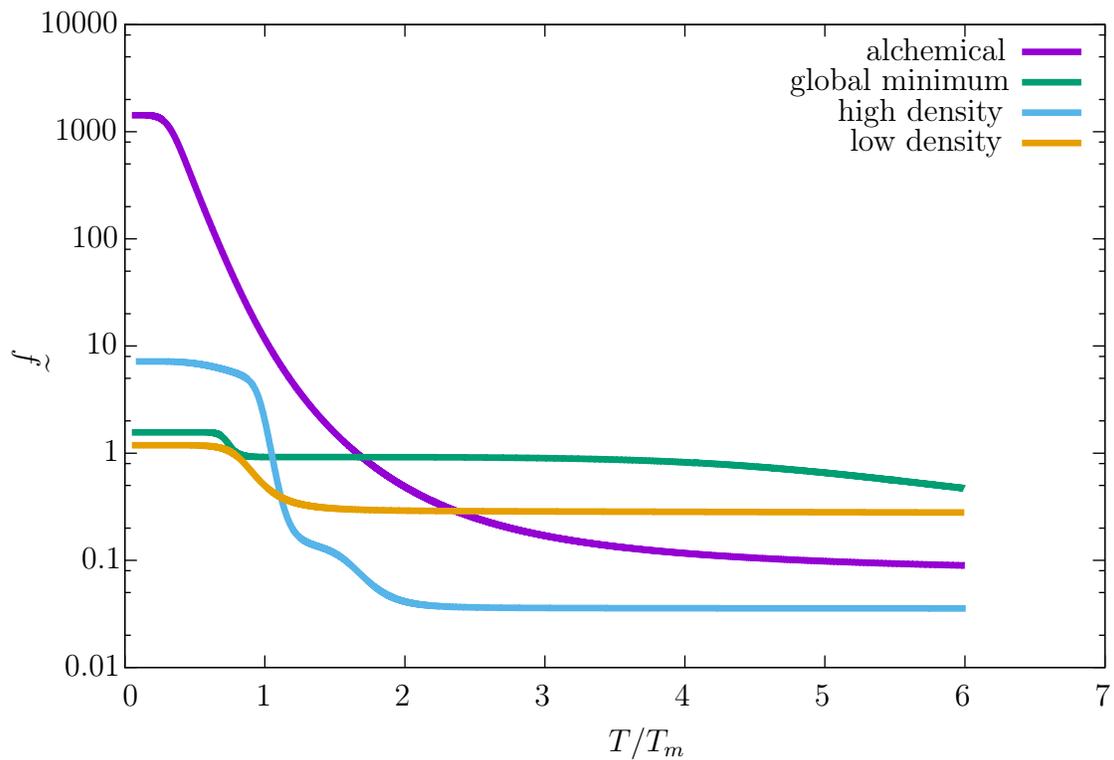}
    \caption{The variation of the frustration indices with temperature for the four databases. The temperature has been
        scaled by the temperature of the greatest value of the heat capacity, $T_{m}$,
        for each database.}
    \label{fig:frust}
\end{figure}

To compare the landscapes more quantitatively, we have calculated the frustration
indices for the four systems, shown in figure \ref{fig:frust}. To facilitate comparison,
the temperature has been scaled for each system by $T_{m}$, the value at which
the heat capacity has a maximum. Heat capacities were calculated within the harmonic
superposition approximation.\cite{Wales93} A mass must be assigned for each degree
of freedom and it was not obvious what mass to use for the alchemical variables. Unit
mass was arbitrarily chosen, as for the atomic coordinates. Therefore,
care should be taken to not overinterpret the value of the frustration index, but we
can make some general observations, which were found to be robust to variation
of the fictitious mass.

The alchemical database is very frustrated at low temperatures, indicating
relatively high barriers between the global minimum and other minima with significant occupation probabilities.
The energies and alchemical parameters of the ten lowest energy alchemical minima
are shown in table \ref{tab:alchmin}, demonstrating that the low energy minima
are widely separated in alchemical space. The pathways between these minima must
cross unfavourable alchemical regions, giving high barriers.
The other three systems exhibit lower frustration. At high temperatures, the 
frustration disappears for the alchemical system as the barriers become thermally accessible.
Due to the higher frustration, we expect self-assembly on this alchemical
landscape to be more difficult than on the fixed parameter landscapes, especially 
at low temperature. The system would become trapped in local minima with thermally
inaccessible barriers preventing progress towards the global minimum.

\begin{table}
    \begin{tabular}{ccc}
        Energy $/ \epsilon$& $k$ & $\phi$ \\ \hline
        $-4.9693$ & 8.74  & 4.37 \\ % Min 157
        $-4.9690$ & 17.42 & 5.60 \\ % Min 252
        $-4.7570$ & 13.40 & 4.96 \\ % Min 419
        $-4.6784$ & 15.38 & 5.24 \\ % Min 138
        $-4.5811$ & 8.57  & 4.33 \\ % Min 41
        $-4.5798$ & 17.02 & 5.50 \\ % Min 38
        $-4.5518$ & 15.45 & 5.14 \\ % Min 513
        $-4.4844$ & 17.43 & 5.40 \\ % Min 276
        $-4.4436$ & 16.20 & 5.26 \\ % Min 231
        $-4.4314$ & 18.19 & 5.32 \\ % Min 6
    \end{tabular}
    \caption{The energies, $k$ and $\phi$ values of the ten lowest energy alchemical
             minima. The minima are widely separated in alchemical space.}
    \label{tab:alchmin}
\end{table}

All stationary points on the alchemical landscape must be stationary points on
the corresponding fixed parameter landscape when the parameters are equal to those of the alchemical
minimum. The inverse is not necessarily true: a stationary point on a fixed landscape
is unlikely to remain a stationary point when the parameters are allowed to vary.
An alchemical minimum must be a fixed parameter minimum, but an alchemical
transition state could remain a transition state or become a minimum once the
parameters are fixed. The second case indicates that the reaction coordinate
at the transition state has significant components for the alchemical degrees of freedom.
This situation is analogous to a martensitic transition, in which the reaction coordinate
involves a change in lattice parameters rather than particle coordinates.\cite{PatelC53}

Of the 10386 alchemical transition states in the database, 346 became minima
when the parameters were fixed. Although it may seem sensible to call these `pure
alchemical transitions', the reaction coordinate may not be precisely localised on
the alchemical degrees of freedom. We can quantify the `alchemical
content' of a transition state by taking the dot product of the reaction coordinate
with a normalised vector parallel to the alchemical degrees of freedom. The reaction
coordinate is the normalised eigenvector corresponding to the single negative
Hessian eigenvalue at the transition state.\cite{MurrellL68}
In table \ref{tab:alch_content} we present the root mean square (RMS) $k$ and $\phi$
content for both classes of alchemical transition state. While it is clear that those
alchemical transition states that become minima tend to have a much greater alchemical
component in the reaction coordinate, and that $k$ tends to have a greater
contribution than $\phi$, they are not purely alchemical. Conversely,
those transition states that remain as transition states tend to have a very small
alchemical component and are almost purely described in terms of Cartesian coordinate
displacements.

\begin{table}
    \begin{tabular}{ccc}
        Transition state becomes... & $k$ contribution & $\phi$ contribution \\ \hline
        Transition state & 0.0396 & 0.019 \\
        Minimum & 0.497 & 0.154 \\
    \end{tabular}
    \caption{RMS $k$ and $\phi$ contributions to the reaction coordinate at
        alchemical transition states, for those transition states that become
        minima when the parameters are fixed, and for those that remain transition
        states.}
    \label{tab:alch_content}
\end{table}

\section{Conclusions}
We have explored the potential energy landscape of small clusters in an extended system where
the parameters of the potential are also variables. As proof of concept, we
have shown that geometrical approaches for exploring the landscape are feasible and
allow rapid location of the global minimum.

We have compared some properties of the alchemical system to systems with fixed
parameters, showing that there is a significant difference between introducing
alchemical degrees of freedom compared to adding another particle.
Although in this particular case, it seems that high frustration on the alchemical landscape
would hinder self-assembly in
an experimental system, that is no barrier to basin-hopping global optimisation
and theoretical prediction of the best parameters is possible.

We also analyse characteristics of pathways on the alchemical surface,
showing that some are almost purely describable in terms of atomic displacements,
while others contain a significant amount of alchemical character.

We expect that digital alchemy will become an important theoretical tool, as it
has general applicability to a wide range of systems. In addition to ongoing work
on the shape of colloids, we envision coarse-grained potentials with variable
parameters, `floppy-bodies' - previously rigid bodies allowed to change shape
in a restricted manner - and protein mutations.

\subsection*{Conflicts of Interest}
There are no conflicts of interest to declare.

\begin{acknowledgement}
The authors thank James Proctor and Pengji Zhou for comments on the original manuscript.
Financial support was provided by the United Kingdom Engineering and Physical
Sciences Research Council (EPSRC).
Data may be accessed at <http://doi.org/10.5281/zenodo.3256234>
\end{acknowledgement}

\newpage

\FloatBarrier

\def\actam{Acta Metall.\xspace}
\def\acp{Adv. Chem. Phys.\xspace}
\def\arpc{Annu. Rev. Phys. Chem.\xspace}
\def\acsnano{ACS Nano\xspace}
\def\compphyscomm{Comput. Phys. Commun.\xspace}
\def\cpl{Chem. Phys. Lett.\xspace}
\def\cosis{Curr. Opin. Colloid Interface Sci.\xspace}
\def\jcompphys{J. Comput. Phys.\xspace}
\def\jcp{J. Chem. Phys.\xspace}
\def\jmstt{J. Mol. Struc-Theochem\xspace}
\def\jpca{J. Phys. Chem. A\xspace}
\def\macrorc{Macromol. Rapid Commun.\xspace}
\def\molp{Mol. Phys.\xspace}
\def\nanos{Nanoscale\xspace}
\def\nanol{Nano Lett.\xspace}
\def\nat{Nature\xspace}
\def\natm{Nat. Mater.\xspace}
\def\tfs{Trans. Faraday Soc.\xspace}
\def\pnas{Proc. Natl. Acad. Sci. USA\xspace}
\def\pra{Phys. Rev. A\xspace}
\def\prb{Phys. Rev. B\xspace}
\def\prot{Proteins: Struct., Funct., Genet.\xspace}
\def\prb{Phys. Rev. B\xspace}
\def\prl{Phys. Rev. Lett.\xspace}
\def\sci{Science\xspace}
\def\softmat{Soft Matter\xspace}
\def\surfsci{Surf. Sci.\xspace}
\bibliography{opp.bib}

\newpage

\FloatBarrier

\begin{tocentry}
    \centering
    \psfrag{k1}{$k = 19.75$}
    \psfrag{phi1}{$\phi = 5$}
    \psfrag{k2}{$k = 6.00$}
    \psfrag{phi2}{$\phi = 1$}
    \includegraphics[width=0.8\textwidth , keepaspectratio=True]{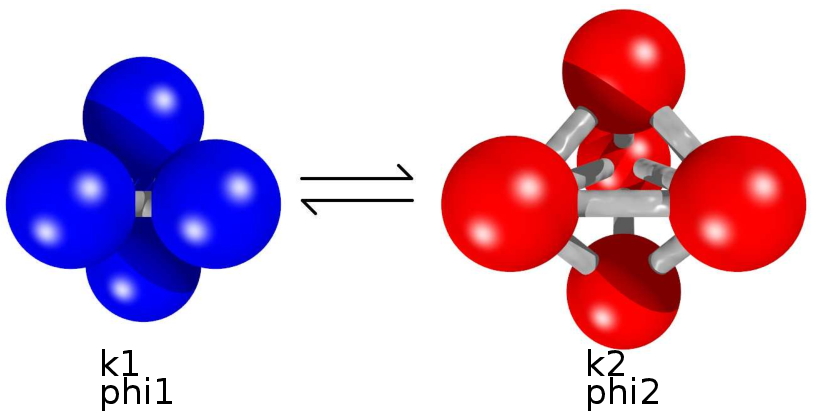}
\end{tocentry}

\end{document}